\documentclass[amsmath,amssymb,aps,reprint,groupedaddress]{revtex4-2}

\usepackage{graphicx}
\usepackage{epstopdf}
\usepackage{dcolumn}
\usepackage{bm}
\usepackage{changes}
\usepackage{multirow}
\usepackage{booktabs}
\usepackage{threeparttable}
\usepackage{algorithmic}
\usepackage{hyperref}
\usepackage{nameref}
\usepackage{xcolor}
\usepackage{soul}
\sethlcolor{red}
\begin{document}

\title{Automatic Extraction and Compensation of P-Bit Device Variations in Large Array Utilizing Boltzmann Machine Training}



\author{Bolin Zhang,$^{1,2}$ Yu Liu,$^{3}$ Tianqi Gao,$^{1,2}$ Jialiang Yin,$^{1,2}$ Zhenyu Guan,$^{3}$ \\Deming Zhang,$^{1,2,}$}
\email{deming.zhang@buaa.edu.cn}
\author{Lang Zeng$^{1,2,}$}
\email{zenglang@buaa.edu.cn}
\affiliation{$^{1}$Fert Beijing Institute, MIIT Key Laboratory of Spintronics, School of Integrated Circuit Science and Engineering, Beihang University, Beijing 100191, China}
\affiliation{$^{2}$National Key Lab of Spintronics, International Innovation Institute, Beihang University, Hangzhou 311115, China}
\affiliation{$^{3}$School of Cyber Science and Technology, Beihang University, Beijing 100191, China.}


\date{\today}

\begin{abstract}
Probabilistic Bit (P-Bit) device serves as the core hardware for implementing Ising computation. However, the severe intrinsic variations of stochastic P-Bit devices hinder the large-scale expansion of the P-Bit array, significantly limiting the practical usage of Ising computation. In this work, a behavioral model which attributes P-Bit variations to two parameters $\alpha$ and $\Delta V$ is proposed. Then the weight compensation method is introduced, which can mitigate $\alpha$ and $\Delta V$ of P-Bits device variations by rederiving the weight matrix, enabling them to compute as ideal identical P-Bits without the need for weights retraining. Accurately extracting the $\alpha$ and $\Delta V$ simultaneously from a large P-Bit array which is prerequisite for the weight compensation method is a crucial and challenging task. To solve this obstacle, we present the novel automatic variation extraction algorithm which can extract device variations of each P-Bit in a large array based on Boltzmann machine learning. In order for the accurate extraction of variations from an extendable P-Bit array, an Ising Hamiltonian based on 3D ferromagnetic model is constructed, achieving precise and scalable array variation extraction. The proposed Automatic Extraction and Compensation algorithm is utilized to solve both 16-city traveling salesman problem(TSP) and 21-bit integer factorization on a large P-Bit array with variation, demonstrating its accuracy, transferability, and scalability.
\end{abstract}



\maketitle
\section{Introduction}\label{sec1}
The traditional computing systems based on the von Neumann architecture typically utilize deterministic binary bits to encode information and perform calculations, leading to inefficient solutions for combinatorial optimization problems, which are often NP-hard or NP-complete problems, such as the knapsack problem, integer factorization, and traveling salesman problem (TSP). Probabilistic Bit (P-Bit), as the core device to construct the hardware entity of Ising computation~\cite{Datta2}, owing to its inherent stochasticity, is proposed for efficiently solving combinatorial optimization problems. Several hardware implementations of P-Bit devices are proposed, for instance, stochastic low barrier magnetic tunneling junctions (MTJ)~\cite{Datta1,Datta3,Datta4,Datta5,Datta6}, spin orbit torque (SOT) driven MTJ ~\cite{IEDM2}, non-linear oscillators~\cite{SHIL1,SHIL2,SHIL3,SHIL4,SHIL5,Zeng2}, diffusive memristors~\cite{Kr1}, metal-insulator transition based P-Bit~\cite{Kr2}, optical parametric oscillators~\cite{optical1,optical2,optical3,optical4,optical5}, CMOS circuits~\cite{CMOS1,CMOS2,CMOS3,CMOS4,CMOS5}, and resistive random access memories (RRAMs)~\cite{RRAM1}.

Among them, MTJ P-Bit devices offer advantages like high speed ($\sim$$ns$), ultra-low power ($\sim$$\mu W$) and small footprint ($\sim$$10nm$). However, in order to implement the stochastic property of P-Bit, the energy barrier $\Delta E$ of MTJ P-Bit need to be reduced to approximately $1k_BT$. In practice, it is very hard to fabricate multiple MTJ P-Bits with the same $\Delta E$, especially the $\Delta E$ is too small to be controlled. This leads to severe intrinsic variation in MTJ P-Bit devices~\cite{Lowba1,Lowba2}. To ensure the correctness of Ising computation, the P-Bit devices are required to be calibrated one by one, which is prohibitively expensive for large P-Bit array. Evenworse, within an analog P-Bit system, the individual variations of P-Bits remain elusive to direct measurement and calibration. Several methods have been presented to address the variation issue, including time division multiplexing (TDM)~\cite{IEDM1,Zeng1}, applying external magnetic fields and voltages~\cite{Variation1}, and configuring the suitable resistances~\cite{Datta1}. However, these approaches are either operationally complex or not scalable.

Boltzmann machines are stochastic recurrent neural networks inspired by Boltzmann distribution and have widespread applications in generative machine learning~\cite{Gml1,Gml2}. Recently, hardware implementations of restricted Boltzmann machines have been proposed~\cite{RBM1,RBM2}. In response to the variability of MTJ P-Bit devices, an in-situ learning approach based on fully visible Boltzmann machine (FVBM) has been introduced~\cite{Alg1}, which updates the weight matrix to adapt to P-Bit variation. This approach regards P-Bits with variations as a unified system and trains suitable weights at the system level, rather than calibration. However, it requires retraining weight matrix for different problem instances, significantly affecting its scalability and increasing the complexity.

The paper is structured as follows: Section~\ref{sec2} presents a behavioral model of P-Bits incorporating $\alpha$ and $\Delta V$. Subsequently, by analyzing computations of the AND gate using both ideal and real P-Bits, it underscores the pressing need to address P-Bit variation. A weight compensation algorithm is then proposed in order to nullify variation by rederiving the weight matrix. Central to this methodology is the imperative requirement for precise knowledge of the $\alpha$ and $\Delta V$ values for each P-Bit. Consequently, Section~\ref{sec3} introduces a variation extraction algorithm based on the Boltzmann machine to capture individual P-Bit variations. Examination of the extraction inaccuracies on the AND gate array reveals the necessity of constructing a suitable density matrix, leading to the proposal of an adaptable 3D ferromagnetic model capable of achieving accurate and scalable array variation extraction. In Section~\ref{sec4}, we demonstrate the effectiveness, transferability, and scalability of our approach by successfully solving the 16-city TSP and 21-bit integer factorization on a large P-Bit array with corrected variations given by the Automatic Extraction and Compensation methods.
\section{Weight Matrix Rederivation to Compensate P-Bit Array Variation}\label{sec2}
 P-Bit device is the main building block of the Ising computer. It is actually a binary stochastic neuron (BSN)~\cite{BSN1,BSN2}, which can fluctuate between -1 and 1 with probability that is tuned by the input voltage (or current). The curve of controlled probability with respect to input voltage is usually Sigmoid-like. The behavioral model of P-Bit is written as~\cite{Alg3}:
\begin{equation}
m(t+1)=sgn(tanh(V(t))-r(t))
\label{beha1}
\end{equation}
where $m(t+1)$ is the output P-Bit state at the next time step, $V(t)$ represents the input voltage, $tanh$ implements the Sigmoid function, and $r(t)$ is a uniformly distributed random number between -1 and 1.

\subsection{Detrimental Impact of P-Bit Array Device Variation on Ising Computation}\label{sec2a}
\begin{figure}[!tb]
\centerline{\includegraphics[width=1.0\columnwidth]{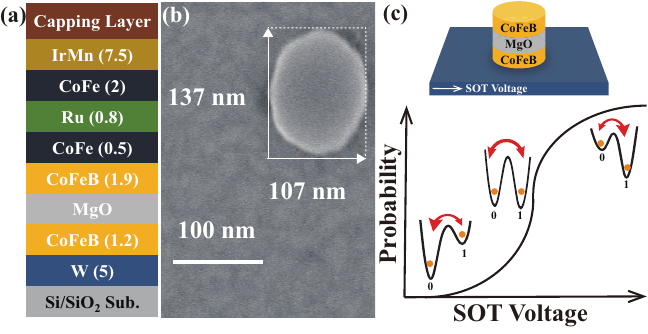}}
\caption{The general overview of fabricated SOT P-Bit device. (a)The deposited structure of stack. (b)The SEM image of SOT MTJ. (c)The operation mechanism of SOT P-Bit. The double-wall energy barrier can be modulated by SOT voltage.}
\label{fig1_n2}
\end{figure}
\begin{figure}[!tb]
\centerline{\includegraphics[width=0.8\columnwidth]{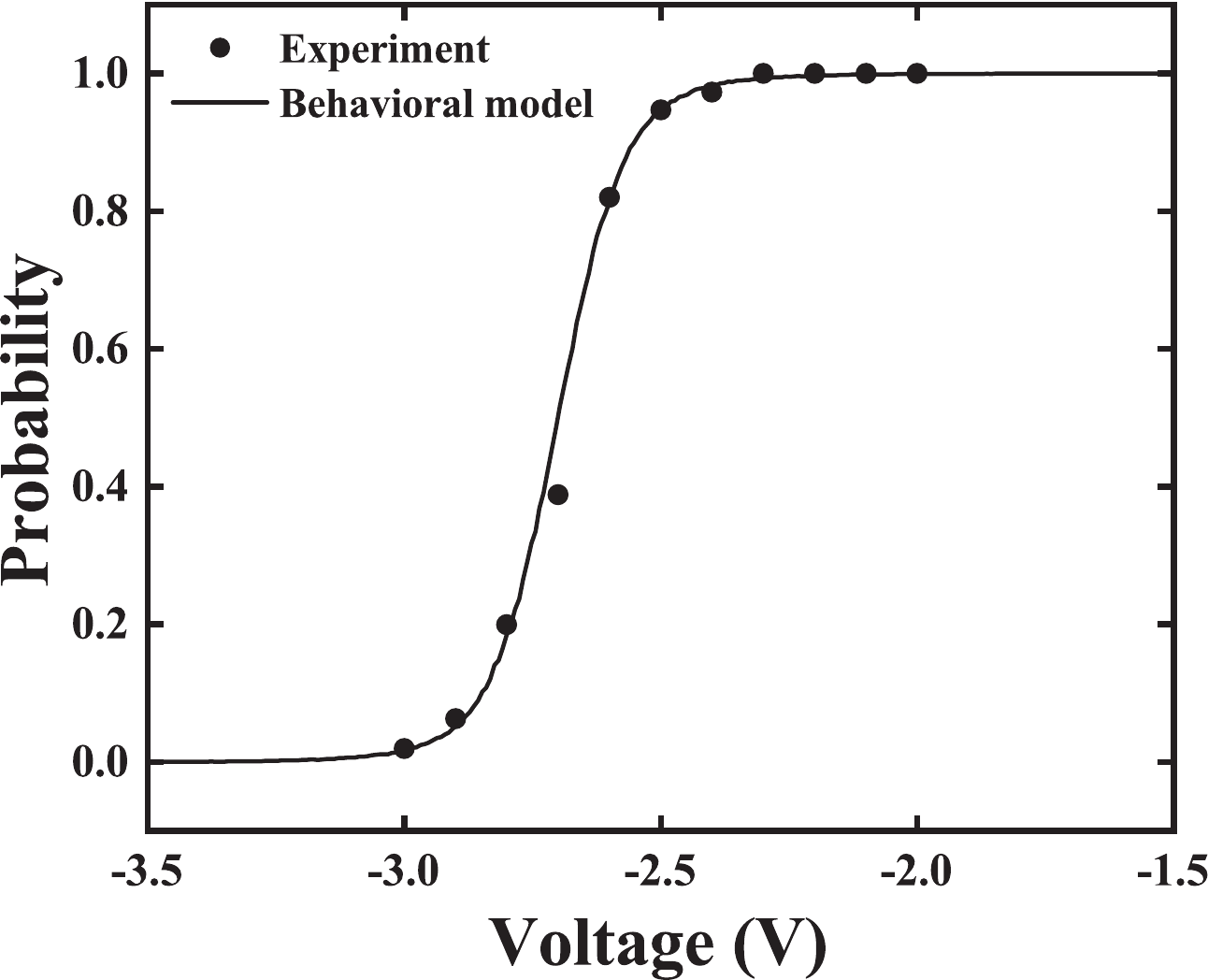}}
\caption{The behavioral model can fit the experimentally measured output probability curve very well.}
\label{fig1_n}
\end{figure}
\begin{figure}[!tb]
\centerline{\includegraphics[width=0.8\columnwidth]{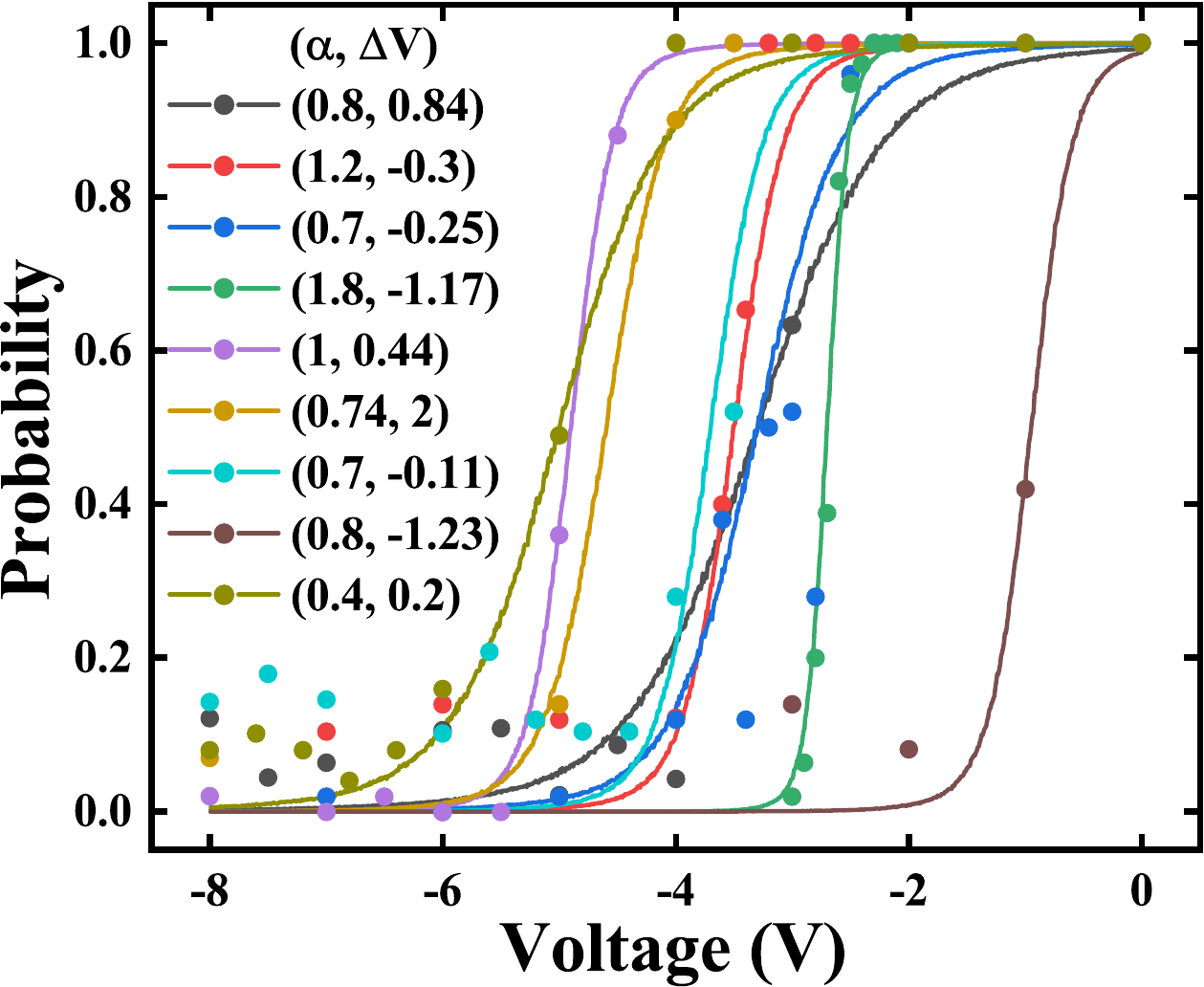}}
\caption{The experimental Sigmoid curves fitted by the modified behavioral model exhibit large intrinsic variation which includes $\alpha$ and $\Delta V$. $\alpha$ describes stretching or compression in shape and $\Delta V$ represents rigid shift.}
\label{fig1}
\end{figure}

 In our previous work, an ultra-fast field-free stochastic SOT P-Bit device is proposed and fabricated~\cite{IEDM2}. As shown in the Fig~\ref{fig1_n2}(a), a stack was deposited consisting of IrMn(7.5)/CoFe(2)/Ru(0.8)/CoFe(2)/CoFeB(1.9)/MgO
 /CoFeB(1.2)/W(5). Fig~\ref{fig1_n2}(b) displayed the SEM image, the stack was etched into elliptical MTJs with a major axis of 137 nm and a minor axis of 107 nm. As illustrated in Fig~\ref{fig1_n2}(c), The SOT MTJ structure involves adding a heavy metal layer beneath an MTJ, which is used to apply a SOT voltage. This voltage can influence the double-wall energy barrier via the SOT effect. The elliptical MTJ has a stray field that inherently favors the ‘0’ state. When the SOT voltage is low, the SOT MTJ tends to stay in the ‘0’ state. As the SOT voltage increases, the probabilities of the ‘0’ and ‘1’ states become 50/50\%. While the SOT voltage is sufficiently high, the SOT MTJ is more likely to remain in the ‘1’ state. The Sigmoid curve of SOT P-Bit device is fitted by the behavioral model, as shown in Fig~\ref{fig1_n}. The behavioral model can fit the experimentally measured data well. However, upon measuring numerous SOT P-Bit devices, we found that the experimentally measured Sigmoid curves exhibit large intrinsic variation. As displayed in Fig~\ref{fig1}, it is clearly observed that, compared to the ideal Sigmoid curve, the deviations of the real curve fall into two categories: stretched or compressed in shape, and rigid shift. This characteristic is also illustrated in both~\cite{IEDM2} and~\cite{Yang1}. In ref~\cite{IEDM2}, we have proposed to use $\alpha$ to describe the degree of stretching and compression, and $\Delta V$ to describe the rigid shift of the Sigmoid curve. The behavioral model that incorporates $\alpha$ and $\Delta V$ to describe device variations is written as:
\begin{equation}
m(t+1)=sgn(tanh(\alpha V(t)+\Delta V)-r(t))
\label{beha2}
\end{equation}
Based on the modified behavioral model, we can fit our experimental Sigmoid curves with different combinations of $\alpha$ and $\Delta V$. However, these device-to-device variations severely compromise the accuracy of Ising computation. To further examine the effect of the P-Bit device variation on Ising computation quantitatively, we connect three 3 P-Bits to construct an AND logic gate. In Ising computing, the input voltage for each P-Bit will be determined as follows:
\begin{equation}
V_i(t+1)=\beta(\sum_{j}W_{ij}m_j(t)+B_i)
\label{beha3}
\end{equation}
where $\beta$ is the inverse pseudo-temperature, $W_{ij}$ is the coupling weight between $i_{th}$ and $j_{th}$ P-Bit, and $B_i$ is the bias of the $i_{th}$ P-Bit. The input voltage $V_i$ will be calculated from the previous states of other P-Bits, which means the P-Bits are coupled with each other.

\begin{figure}[!tb]
\centerline{\includegraphics[width=0.8\columnwidth]{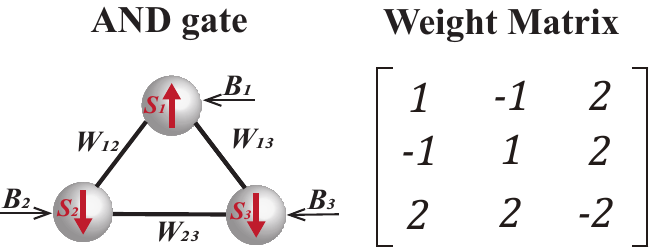}}
\caption{The AND gate comprises three P-Bits, and the weight matrix for the AND gate is provided on the right.}
\label{fig2_n}
\end{figure}
\begin{figure}[!tb]
\centerline{\includegraphics[width=0.8\columnwidth]{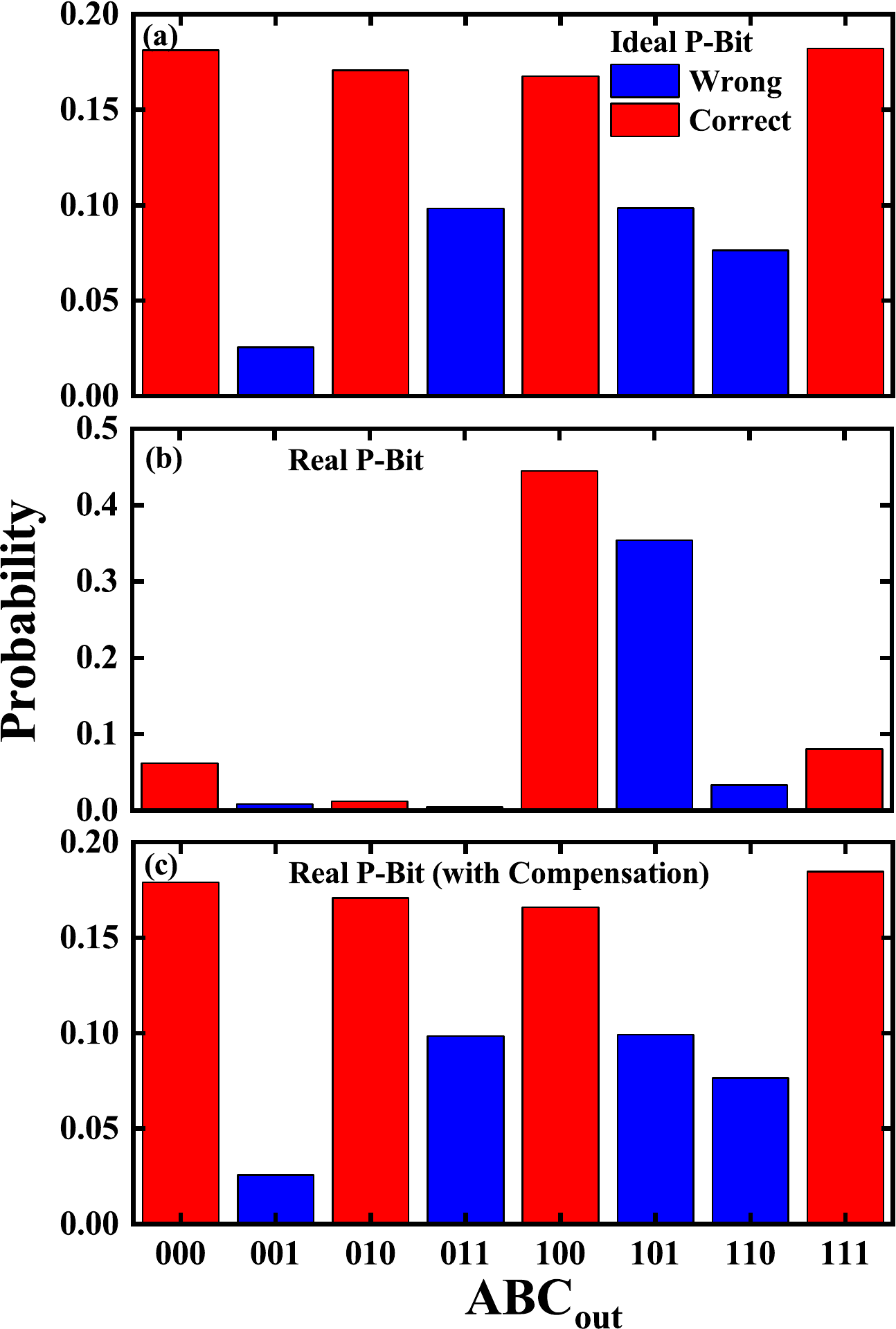}}
\caption{Under three different P-Bit configurations, the computational outcomes of the AND gate are depicted, with red denoting the correct state and blue representing the wrong state. (a) The ideal P-Bit array successfully completes the computation, with an evenly distributed occurrence of the four correct states. (b) The real P-Bit array with variations fails to accurately complete the computation, displaying a significantly low occurrence of three correct states. (c) The real P-Bit array with weight compensation achieves accurate computation akin to the ideal P-Bit array.}
\label{fig2}
\end{figure}
As depicted in Fig.~\ref{fig2_n}, three P-Bits are interconnected. The coupling weight matrix for the AND gate is also provided, and the truth table for the AND gate is given as Table~\ref{AND}. Each truth value of AND gate has a probability: $p=1/4=0.25$.
\begin{table}[b]
\caption{\label{AND}%
Truth table for AND gate
}
\begin{ruledtabular}
\begin{tabular}{lcdr}
\textrm{A}&
\textrm{B}&
\textrm{$C_{out}$}&
\textrm{$P_{ideal}$}\\
\colrule
0 & 0 & 0 & 0.25\\
0 & 1 & 0 & 0.25\\
1 & 0 & 0 & 0.25\\
1 & 1 & 1 & 0.25\\
\end{tabular}
\end{ruledtabular}
\end{table}

The result of AND gate computation in three scenarios of P-Bit variation is shown in Fig.~\ref{fig2}. As displayed in Fig.~\ref{fig2} (a), the ideal P-Bit array obtains accurate results in AND gate computation at $\beta=1$, each of the four correct states is approximately $18\%$ and this make $72\%$ total accuracy. While in Fig.~\ref{fig2}(b), the real P-Bit array extracted from the experimental data with different $\alpha$ and $\Delta V$ can not get the accurate results in Ising computing. The accuracy from real P-Bit array is $59\%$, much smaller than the ideal P-Bit array. It is even worse that the four correct states are not equal, the states of P-Bits in computing are trapped in two states of high probability rather than fluctuating among the four correct states as expected.
\begin{figure}[!tb]
\centerline{\includegraphics[width=0.9\columnwidth]{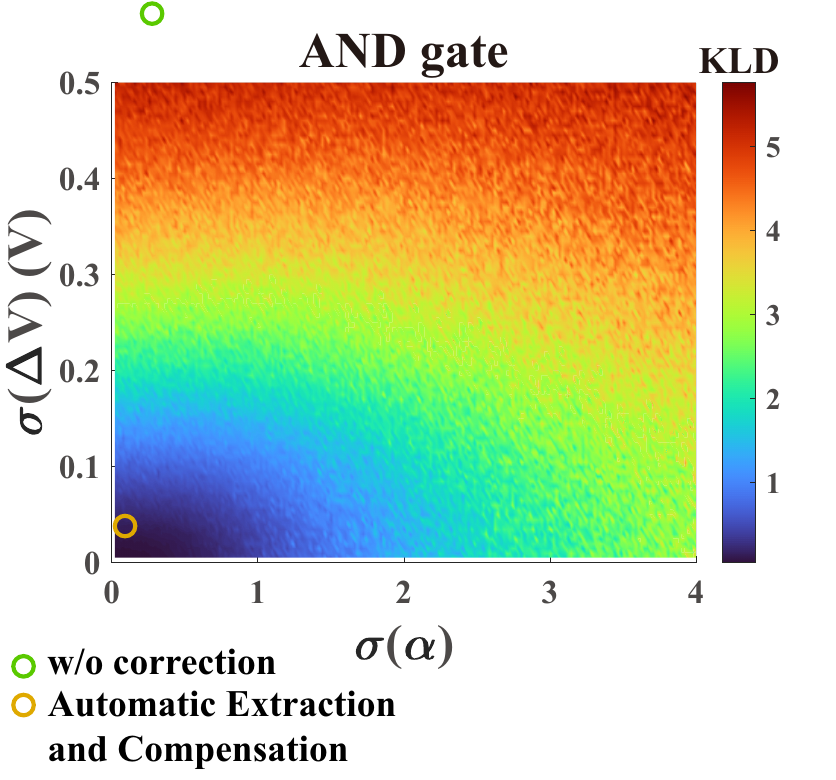}}
\caption{The Monte Carlo analysis of $\alpha$ and $\Delta V$ on the KLD for AND gate. Each pair of different $(\sigma(\alpha), \sigma(\Delta V))$ has been sampled 1000 times, and their KLDs are averaged.}
\label{fig2_mk}
\end{figure}

We employed the Monte Carlo method to perform a statistic analysis of the AND gate. To better evaluate the computational results, we utilized the Kullback-Leibler divergence(KLD)\cite{Kld}:
\begin{equation}
KLD[P_{calc}||P_{des}]=\sum_{m}P_{des}(m)log(P_{des}(m)/P_{calc}(m))
\label{KLD1}
\end{equation}

where the $P_{des}$ is the designed probability distribution, $P_{calc}$ is the probability distribution calculated from Ising computer. The KLD measures the similarity between two probability distributions. A smaller KLD value indicates a better quality of the computed results.
As shown in the Fig.~\ref{fig2_mk}, the horizontal axis represents the standard deviation of $\alpha$, while the vertical axis represents the standard deviation of $\Delta V$. Each pair of different $(\sigma(\alpha), \sigma(\Delta V))$ is sampled 1000 times, and their KLDs are averaged and plotted as one point in the figure. As illustrated in the Fig.~\ref{fig2_mk}, a small standard deviation of $\Delta V$ results in a significantly large KLD for the AND gate computation, while the tolerance for $\alpha$ is higher. When the standard deviation of $\alpha$ is less than 0.5, the computation can ensure sufficient accuracy. Additionally, it has been indicated that the device variation measured from experiments without correction exceeds the range shown in Fig.~\ref{fig2_mk}.

The variations in the real P-Bit array lead to the wrong results in Ising computing. As the scale of solved problem expands, the negative effect of P-Bit array variation on Ising computation will accumulate. The main drawback in hardware implementation of Ising computing is that the P-Bit devices have to be calibrated individually to get the same Sigmoid curves, which is not practical for large array. A novel approach is required for addressing the P-Bit device variation, instead of expensively calibrating one by one.

\subsection{Transferable Compensation for P-Bit Array Variation by Rederiving Weight Matrix}\label{sec2b}
\begin{figure}[!tb]
\centerline{\includegraphics[width=0.9\columnwidth]{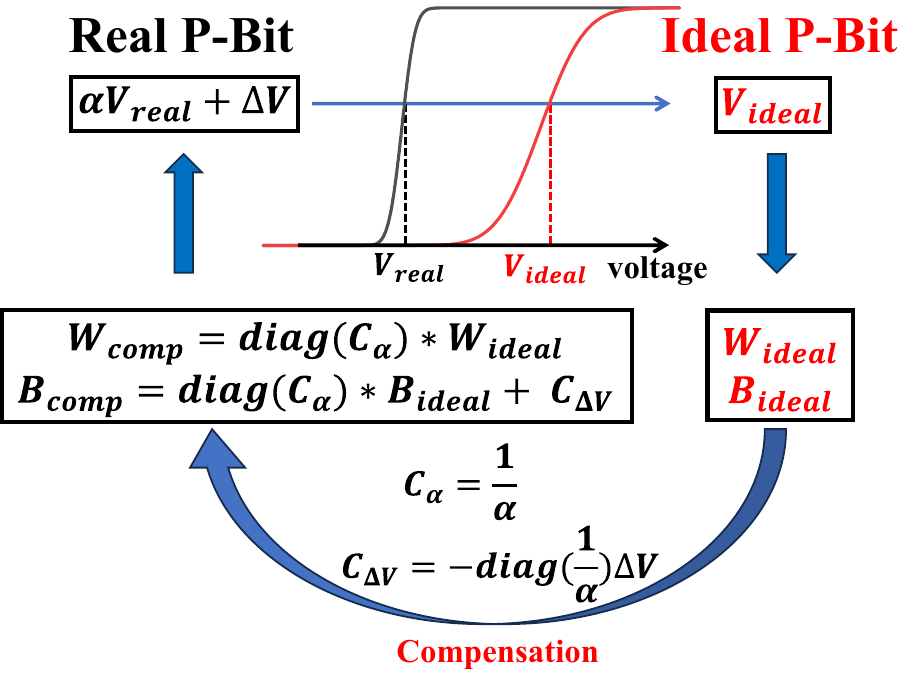}}
\caption{The process of weight matrix compensation. To achieve the same output probability, it is necessary to satisfy that: $\alpha V_{real}+\Delta V=V_{ideal}$. After the derivation, the new weight and bias matrices incorporating compensation parameters $C_\alpha$ and $C_{\Delta V}$ enable the real P-Bits to compute like the ideal P-Bits.}
\label{fig3}
\end{figure}

In this paper, we first propose the weight compensation method for addressing the P-Bit array variation. This method compensates the P-Bit device variation by rederiving weight matrix instead of calibrating each P-Bit device individually. In Fig.~\ref{fig3}, the process of the weight matrix compensation is displayed. For the ideal P-Bit, the weight matrix $W_{ideal}$ is only calculated from Ising Hamiltonian corresponding to the problem. While for the real P-Bit array, it is necessary to integrate extra parameters into the weight matrix to mitigate the variations inherent in the P-Bit device array.

In accordance with Eqs.~(\ref{beha1}) and (\ref{beha2}), achieving identical output probabilities for P-Bit $i$ in both the real P-Bit array and the ideal P-Bit array requires:
\begin{equation}
tanh(\alpha V_{real,i}+\Delta V_i)=tanh(V_{ideal,i})
\label{w-c1}
\end{equation}
therefore the applied input voltage must adhere to the subsequent relationship:
\begin{equation}
\alpha_iV_{real,i}+\Delta V_i=V_{ideal,i}
\label{w-c2}
\end{equation}
through a straightforward transformation, we can obtain:
\begin{equation}
V_{real,i}=\frac{1}{\alpha_i}V_{ideal,i}-\frac{1}{\alpha_i}\Delta V_i
\label{w-c3}
\end{equation}
In the context of a whole P-Bit array, the above equation can be articulated in a matrix representation:
\begin{equation}
V_{real}=diag(\frac{1}{\alpha})V_{ideal}-diag(\frac{1}{\alpha})\Delta V
\label{w-c4}
\end{equation}
where $\alpha$, $\Delta V$ and $V$ are $n\times 1$ vectors, $n$ is the number of P-Bits in computing. The $diag(\frac{1}{\alpha})$ is the diagonal matrix generated from vector $1/\alpha$. The Eq.~(\ref{beha3}) can also be expressed in matrix form:
\begin{equation}
V_{input}=Wm+B
\label{w-c5}
\end{equation}
where the $V$, $m$ and $B$ are $n\times 1$ vectors, $W$ is $n\times n$ weight matrix. $W$ and $B$ are calculated from the Ising Hamiltonian based on the solved problem. Upon the substitution of Eq.~(\ref{w-c5}) into Eq.~(\ref{w-c4}), the resulting expression is as follows:
\begin{eqnarray}
V_{real}=diag(\frac{1}{\alpha})W_{ideal}m+\qquad\qquad\qquad \nonumber\\
diag(\frac{1}{\alpha})B_{ideal}-diag(\frac{1}{\alpha})\Delta V
\label{w-c6}
\end{eqnarray}
therefore, we can derive new $W$ and $B$ for compensating the P-Bit array variation, denoting them as:
\begin{eqnarray}
W_{comp}=diag(\frac{1}{\alpha})W_{ideal}\qquad\qquad\qquad\nonumber \\
B_{comp}=diag(\frac{1}{\alpha})B_{ideal}-diag(\frac{1}{\alpha})\Delta V
\label{w-c7}
\end{eqnarray}
through the aforementioned deductions, we have obtained the rederived weight and bias matrices crucial for ensuring the precise Ising computation of the real P-Bit array. We define two compensation parameters: $C_\alpha=1/\alpha$, $C_{\Delta V}=-diag(\frac{1}{\alpha})\Delta V$. The rederived weight matrix and bias matrix can be written as:
\begin{eqnarray}
W_{comp}=diag(C_{\alpha})W_{ideal}\qquad\quad\nonumber \\
B_{comp}=diag(C_{\alpha})B_{ideal}+C_{\Delta V}
\label{w-c8}
\end{eqnarray}

To verify the correctness of the weight matrix compensation, we initially conduct experimental measurements of the Sigmoid curves for three P-Bits and utilize behavioral model to fit these data, thereby obtaining distinct values of $\alpha$ and $\Delta V$. Then we alter the weight matrix for computing AND gate by integrating $C_\alpha$ and $C_{\Delta V}$. As shown in Fig.~\ref{fig2}(c), with the weight matrix compensation method, the real P-Bit array can calculate the accurate results in the AND gate, and the results are similar to that of ideal P-Bit array without variations, which proves the reliability and rightness of the weight compensation.

It is noteworthy that the compensation parameters $C_\alpha$ and $C_{\Delta V}$ solely relate to P-Bit variations, rather than the weight matrix. This implies that even with changes in the solving problem, these compensation parameters remain unchanged. In essence, it eliminates the necessity to retrain the weight matrix. The prerequisite is merely obtaining the P-Bit variation parameters: $\alpha$ and $\Delta V$, owing to our consideration of these variations as inherent and immutable attributes of P-Bit devices. This demonstrates the transferability of the weight compensation method across diverse problem domains.

It has been certified that the weight compensation enables real P-Bit array computes similarly to ideal P-Bit array without calibrating Sigmoid curve one by one or retraining the weight matrix. However, the most challenging key point lies in how to efficiently extract the $\alpha$ and $\Delta V$ from the large P-Bit array. Although calibration is not required, measuring and fitting the variations of P-Bit array is also complicated tasks. We seek to perform an expeditious and precise extraction of variations within extensive P-Bit array.
\begin{figure}[!tb]
\centerline{\includegraphics[width=0.8\columnwidth]{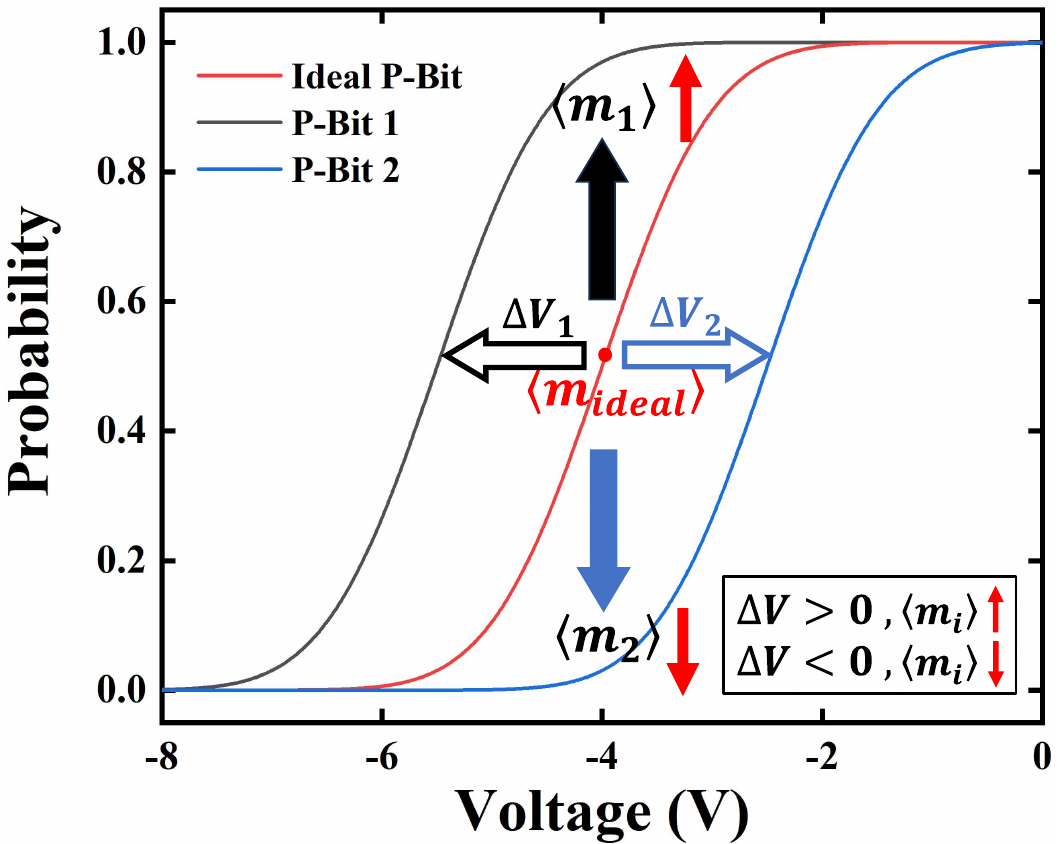}}
\caption{The impact of $\Delta V$ observed on the mean value of P-Bit in Ising computation.}
\label{fig4}
\end{figure}
\begin{figure}[!tb]
\centerline{\includegraphics[width=0.8\columnwidth]{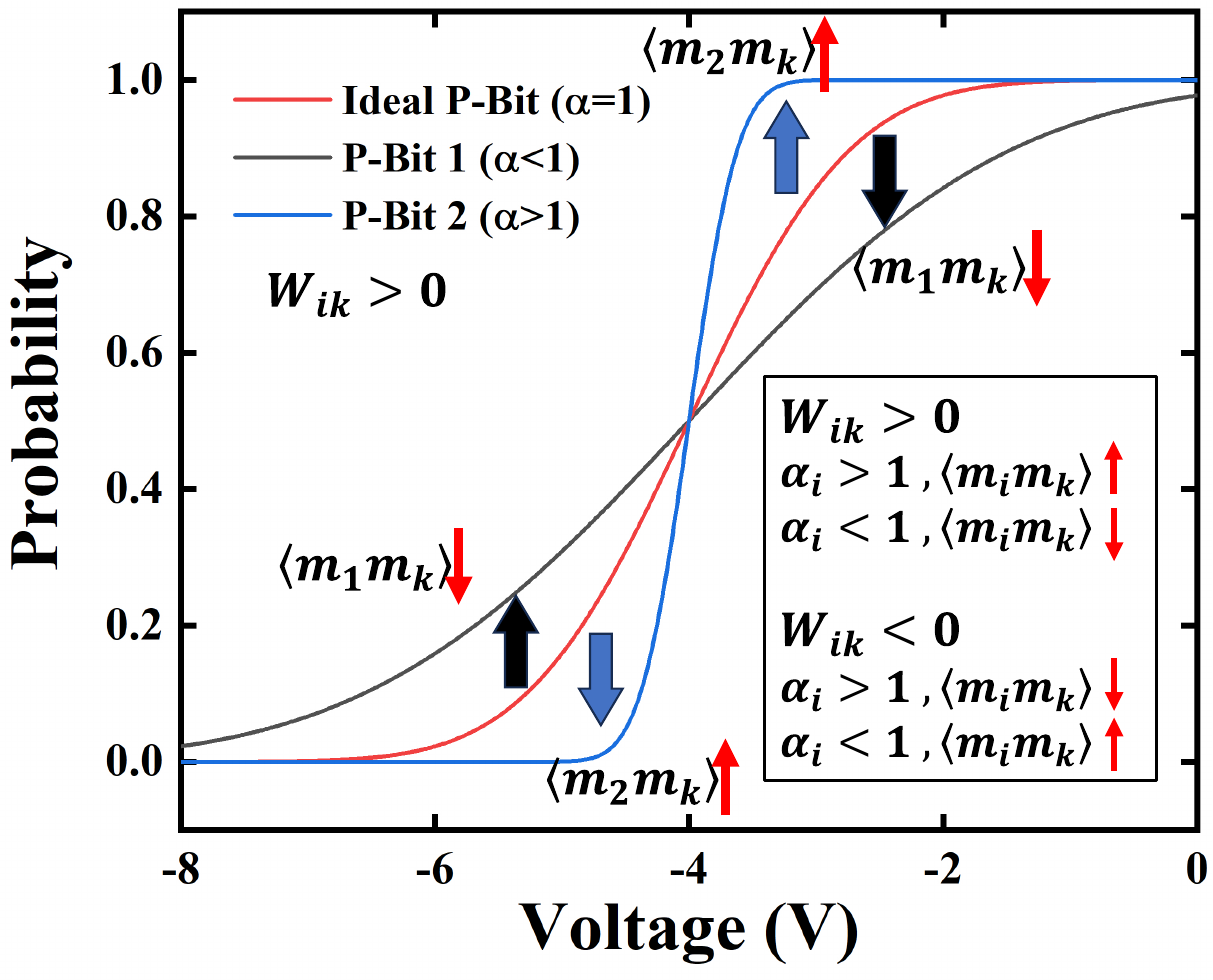}}
\caption{The impact of $\alpha$ observed on the correlation between P-Bits in Ising computation.}
\label{fig5}
\end{figure}
\section{Simultaneous Device Variation Extraction for Whole P-Bit array}\label{sec3}
Boltzmann machine learning plays a significant role in uncovering and analysing, particularly in complex data distributions. Boltzmann machine can be mapped to the hardware system for P-Bit, it trains the weight matrix $W$ for the given data distribution. The learning rule can be written as~\cite{Alg1}:
\begin{equation}
W_{i,j}(t+1)=W_{i,j}(t)+\epsilon (\langle v_iv_j\rangle - \langle m_im_j\rangle-\lambda W_{i,j}(t))
\label{v-e1}
\end{equation}
here $\langle v_iv_j\rangle$ is the average correlation between $i_{th}$ and $j_{th}$ P-Bit in the given data distribution, and $\langle m_im_j\rangle$ is the correlation of the outputs sampled from P-Bit $i$ and P-Bit $j$ in Ising computing, $\lambda$ is the regulation parameter, and $\epsilon$ is the learning rate. In the training process, the $W_{ij}$ will be learned from $\langle m_im_j\rangle$, while the $B_i$ in the weight matrix corresponds to $\langle m_i\rangle$.

The Eq.~(\ref{v-e1}) can only train the required weight matrix for a specific problem. For another problem, it needs to retrain the other weight matrix, which requires more time consumption and is lack of transferability and scalability. However, our objective is to extract the $\alpha$ and $\Delta V$ of each P-Bit device in the real array just once, thinking of them as the inherent attributes of each P-Bit, then use the weight compensation method to solve diverse unconventional problems. Such process is transferable without training again complicatedly.
{\subsection{Automatic Extraction Algorithm Based on Boltzmann Machine Learning}\label{sec3a}
To extract the P-Bit variation by Boltzmann machine learning, we need to observe and analyze the impact of $\alpha$ and $\Delta V$ on the data distributions in Ising computation. Let's treat $\Delta V$ first. As shown in Fig.~\ref{fig4}, the red Sigmoid curve represents the ideal P-Bit, and the black curve is the P-Bit 1, the blue curve is the P-Bit 2. When $\Delta V>0$, the ideal Sigmoid curve rigidly shifts to the left, all points of this curve have also moved upward for the input voltage. Consequently, the $\langle m\rangle$ increases from $\langle m_{ideal}\rangle$ to $\langle m_1\rangle$. Conversely, when $\Delta V<0$, the $\langle m\rangle$ will decrease from $\langle m_{ideal}\rangle$ to $\langle m_2\rangle$. Based on this observation, in instances where the $\langle m\rangle$ is larger than $\langle m_{ideal}\rangle$ during the computational process, it is recommended to diminish the $\Delta V$ in the learning, and vice versa.

As depicted in Fig.~\ref{fig5}, the $\alpha$ of the ideal P-Bit is 1, the Sigmoid curve of P-Bit 1 with $\alpha <1$ is stretched, the Sigmoid curve of P-Bit 2 with $\alpha >1$ is compressed. When $W_{ik}>0$, the interaction between P-Bit $i$ and P-Bit $k$ in Ising computing is positive, then the state of two P-Bits are the same direction. For P-Bit 1, $\alpha <1$, the absolute value of the $\langle m_1\rangle$ have decreased, which results in the decrease of $\langle m_1m_k\rangle$ from $\langle m_1m_k\rangle_{ideal}$. For P-Bit 2, $\alpha >1$, the absolute value of the $\langle m_2\rangle$ have increased, which results in the increase of $\langle m_2m_k\rangle$ from $\langle m_1m_k\rangle_{ideal}$. Building on this insight, in cases where the $\langle m_i m_k \rangle$ is larger than $\langle m_im_k\rangle_{ideal}$, it is advisable to reduce the $\alpha_i$ in the Boltzmann machine learning. And in cases where the $\langle m_i m_k \rangle$ is smaller than $\langle m_im_k\rangle_{ideal}$, $\alpha_i$ should be increased in the Boltzmann machine learning. When $W_{ik}<0$, the above observation will be opposite.

\begin{figure}[!tb]
\centerline{\includegraphics[width=0.9\columnwidth]{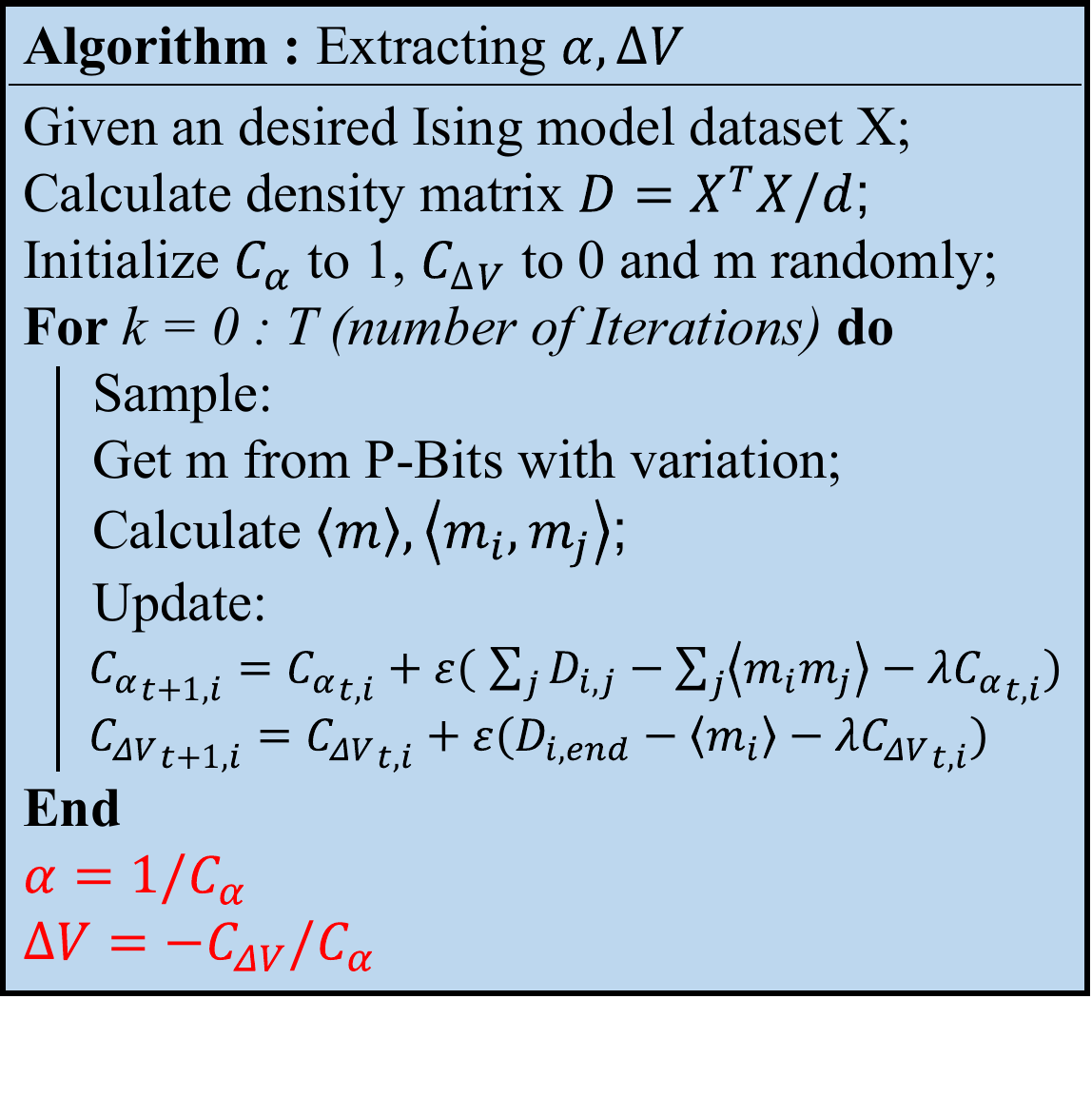}}
\vspace{-20pt}
\caption{Proposed extraction algorithm for P-Bit array variation with given Ising Hamiltonian. It is necessary to initially compute the density matrix $D$ and continuously update the compensation parameters to attain the desired distribution in the sampled Ising computation, ultimately obtaining $\alpha$ and $\Delta V$. The learning rate $\epsilon$ used in this algorithm is 0.0001, while the regulation parameter $\lambda$ is 0.05.}
\label{fig6}
\end{figure}
Based on the analysis above, we propose the extraction algorithm for P-Bit array variation. As shown in Fig.~\ref{fig6}, firstly, the density matrix $D$ is calculated as~\cite{Alg1}:
\begin{equation}
D=X^TX/d
\label{v-e2}
\end{equation}
where $X$ is the dataset corresponding to the problem, $d$ is the order of the matrix. Density matrix $D$ describes the average probability distribution of the correlation between P-Bits. For instance, the $D_{1,2}$ is actually the average correlation $\langle v_1v_2\rangle$ in Eq.~(\ref{v-e1}), which signifies that the desired ideal value of $\langle m_1m_2\rangle$ in Boltzmann machine learning is $D_{1,2}$.

From the perspective of a whole P-Bit array, the $\alpha_i$ will affect the correlation between the P-Bit $i$ and other P-Bits, therefore the compensation parameter $C_{\alpha_i}$ will be learned from the sum of average correlation $\sum_{j} \langle m_im_j\rangle$. While the compensation parameter $C_{\Delta V_i}$ is learned from $\langle m_i\rangle$. After learning of the compensation parameter reach convergence, the variation of each P-Bit device in the array can be extracted as: $\alpha =1/C_\alpha$, $\Delta V=-C_{\Delta V}/C_\alpha$.

\subsection{Inefficiency Extraction based on Traditional Logic Gate}\label{sec3b}
To verify the feasibility of the proposed extraction algorithm, we perform the extraction of the P-Bit array variations on the AND gate. Initially, we can calculate the density matrix $D$ for the AND gate based on the truth table in Table~\ref{AND} and the Eq.~(\ref{v-e2}).
\begin{equation}
D_{AND}=
\begin{bmatrix}
1 & 0 & 0.5 & 0\\
0 & 1 & 0.5 & 0\\
0.5 & 0.5 & 1 & -0.5\\
0 & 0 & -0.5 & 1 \nonumber
\end{bmatrix}
\end{equation}

The density matrix $D$ describes the ideal data distribution for AND gate. In the learning of $\alpha $ and $\Delta V$, our aim is to align the sampled data distribution in Ising computing with the ideal distribution by updating the compensation parameters $C_\alpha$ and $C_{\Delta V}$. Finally, we can obtain the $\alpha$ and $\Delta V$.
\begin{figure}[!tb]
\centerline{\includegraphics[width=0.8\columnwidth]{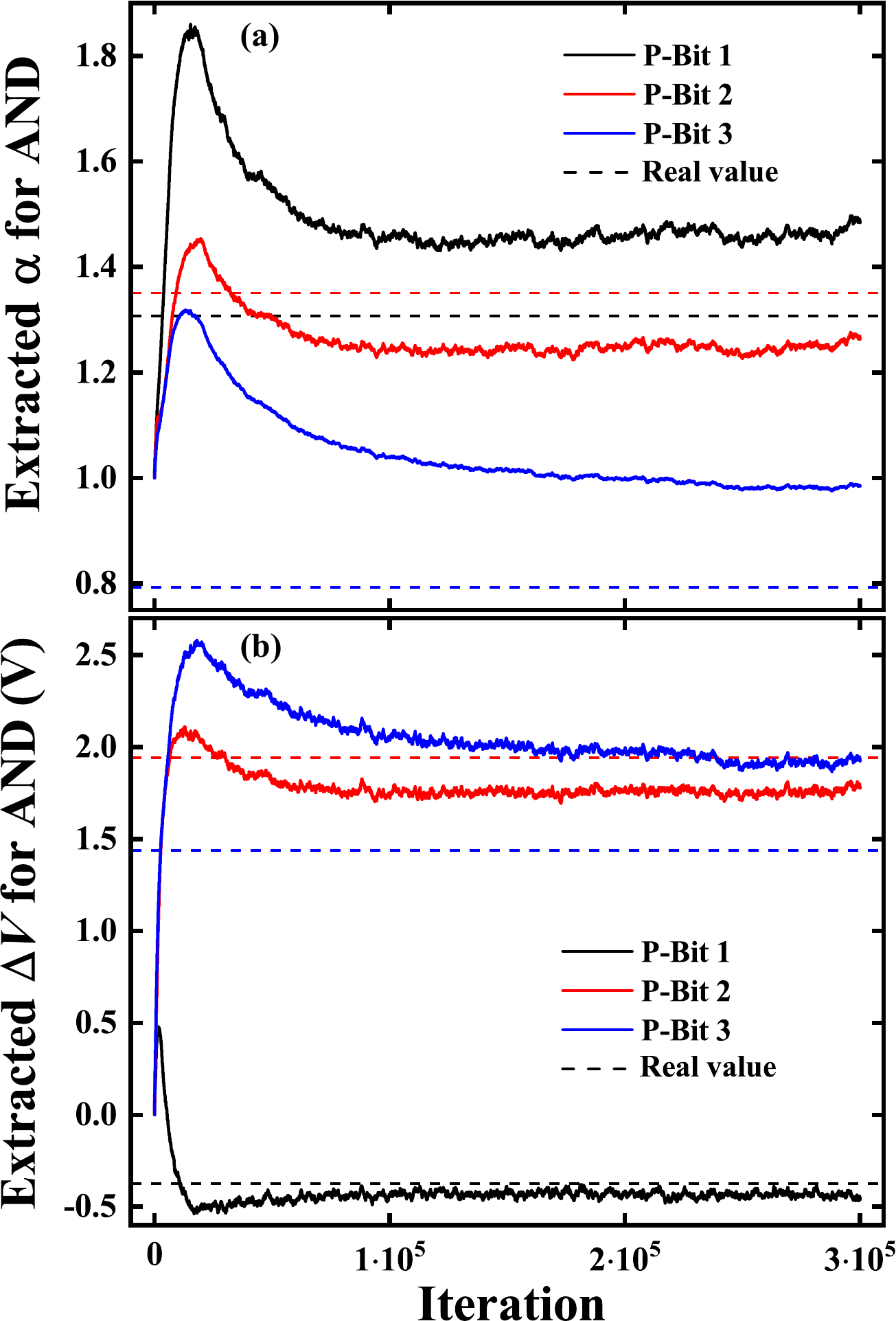}}
\caption{The learning results of presented extraction algorithm on AND gate. The solid lines display the respective extracted value of P-Bits, while the dash lines represent the real value.}
\label{fig7}
\end{figure}

The Fig.~\ref{fig7} displays the outcomes of our proposed extraction algorithm applied to an AND gate. We randomly choose three fabricated SOT P-Bits of different variations. As depicted in Fig.~\ref{fig7}(a), $\alpha_i$ corresponds to P-Bit $i$, the dash line means the real value of P-Bit variation. The training trends of $\alpha_1$ and $\alpha_2$ are correct, and the extracted values of $\alpha_1$ and $\alpha_2$ are somewhat close to their respective real values. However, the extracted $\alpha_3$ is inaccurate. The extracted $\Delta V$ results of P-Bit 1 and P-Bit 2 shown in Fig.~\ref{fig7}(b) are relatively aligned with the real value, while the inaccuracy of extracted $\alpha_3$ has detrimental impact on the $\Delta V_3$, causing the extracted $\Delta V_3$ slightly deviates from the real value. This can be explained that the $\Delta V_3$ is calculated as: $\Delta V_3=-C_{\Delta V_3}/C_{\alpha_3}$.

The proposed extraction algorithm has demonstrated effectiveness in the AND gate. However, it has not yet reached a high level of accuracy. The performance of extracted $\alpha$ is subpar, which significantly influences the training process for $\Delta V$. We shall carefully examine the reasons for the inaccurate training of $\alpha$ in the AND gate.

In the training rule of $\alpha$ displayed in Fig.~\ref{fig6}, $\alpha_i$ is learned from all $\langle m_im_j\rangle$ in the $i_{th}$ row of density matrix. Nevertheless, when $W_{ij}$ in weight matrix comprises both positive and negative values, the training of $\alpha_i$ will be affected, this occurs due to the contrasting training directions of $W_{ij}>0$ and $W_{ij}<0$ as analyzed in Fig.~\ref{fig5}.

Based on the preceding analysis, the weight matrix from the Ising Hamiltonian will notably influence the extraction of $\alpha$ and $\Delta V$. As a consequence, a suitable Ising Hamiltonian for the proposed extraction algorithm should be meticulously constructed. Firstly, the Hamiltonian we construct needs to be scalable for extending the variation extraction algorithm to larger P-Bit arrays. Secondly, the corresponding density matrix of this Hamiltonian should be sparse to facilitate simpler and more accurate extraction. From the density matrix of the AND gate, we observe reduced accuracy in extraction when the density matrix is uneven. Hence, finally, a homogeneous density matrix is essential, showcasing advantages in training convergence rate.

\subsection{3D Ferromagnetic Hamiltonian Model For Scalable Large P-Bit Array}\label{sec3c}
The Ising model, renowned for its depiction of ferromagnetism, prompts a pertinent inquiry: can this ferromagnetic model be harnessed for variation extraction purposes? Drawing inspiration from this, we have formulated a succinct and scalable ferromagnetic model. Within this constructed regular ferromagnetic lattice, spins exclusively interact with their immediate neighboring spins, ensuring the overall sparsity of the system. Notably, the density matrix associated with our developed ferromagnetic model adheres to uniformity required in the aforementioned criteria.

\begin{figure}[!tb]
\centerline{\includegraphics[width=0.98\columnwidth]{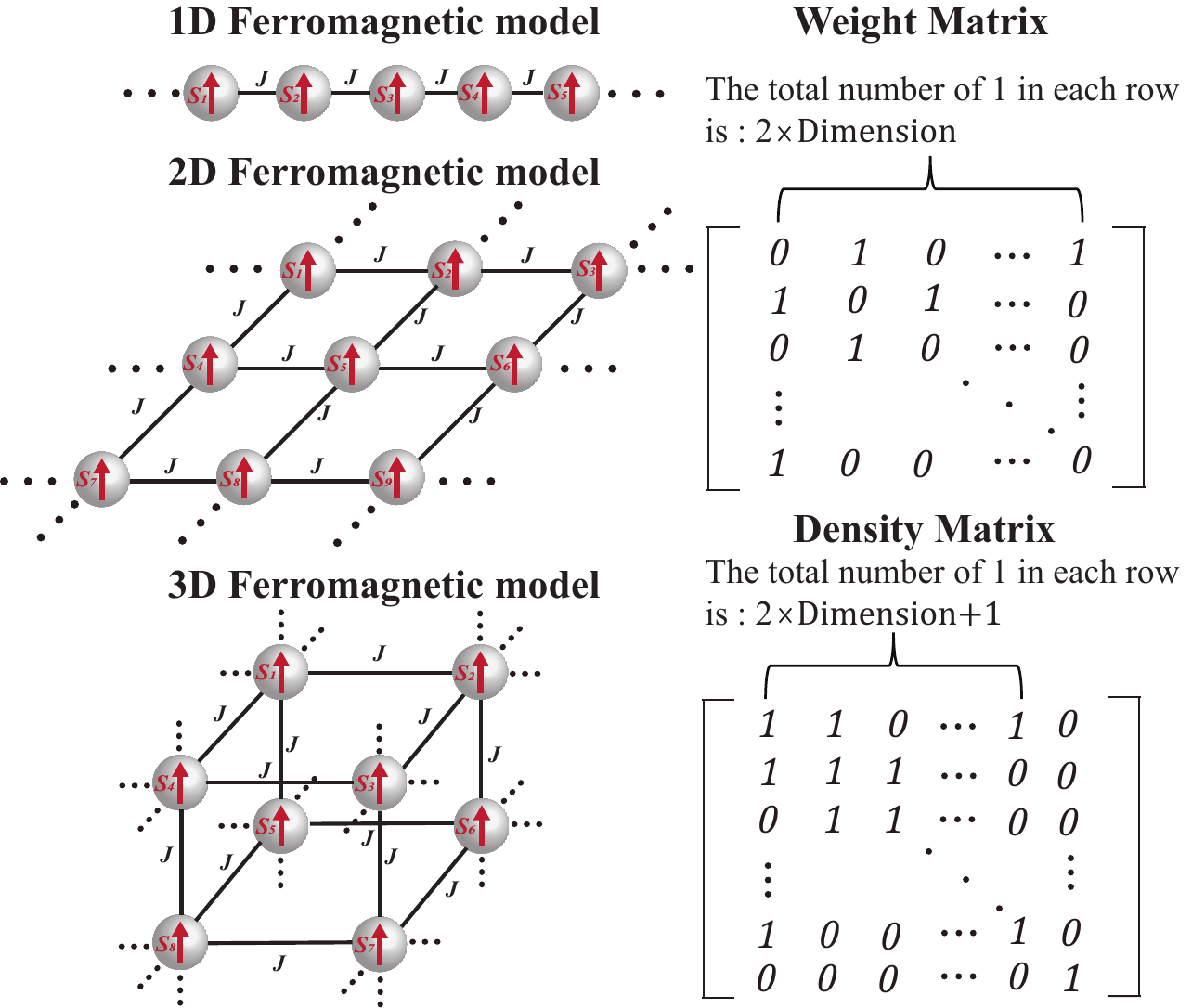}}
\caption{Constructed 1D, 2D and 3D ferromagnetic Hamiltonian model for the extraction algorithm. Their uniform weight matrix and density matrix are provided on the right.}
\label{fig8}
\end{figure}
\begin{figure*}[!tb]
\centerline{\includegraphics[width=2\columnwidth]{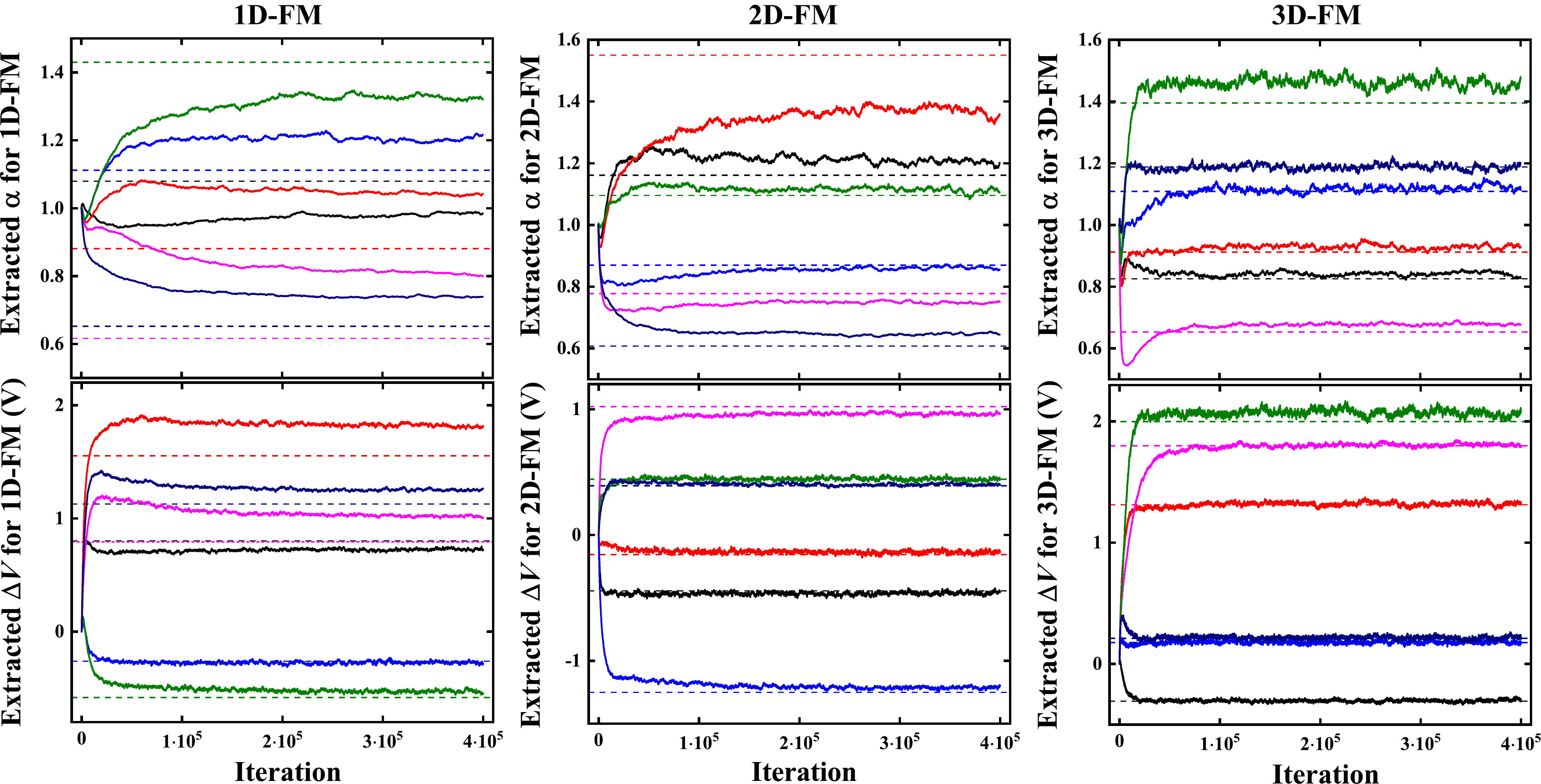}}
\caption{The learning results of extraction algorithm employed on constructed 1D, 2D and 3D ferromagnetic model. As the dimension of ferromagnetic model increases, the accuracy of variation extraction gradually improves. The 3D-FM achieves flawless extraction, with the extracted values perfectly aligning with the real values.}
\label{fig9}
\end{figure*}
\begin{figure}[!tb]
\centerline{\includegraphics[width=0.9\columnwidth]{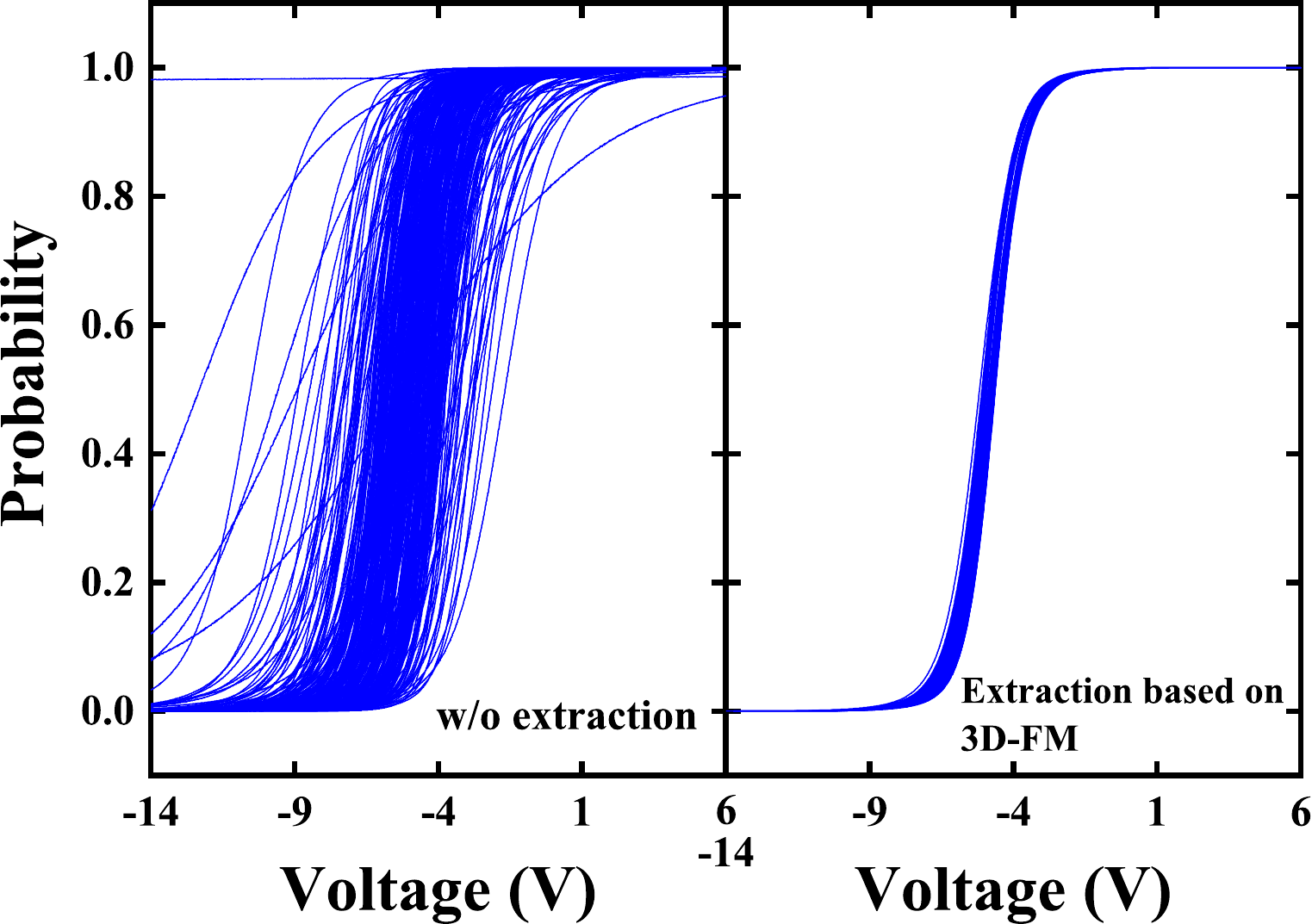}}
\caption{The Sigmoid curves obtained from P-Bit arrays w/o correction and with extraction. According to Table.\ref{STD}, the $(\sigma_\alpha,\sigma_{\Delta V})$ of P-Bit array without extraction is $(0.3,1V)$, while that of P-Bit array with extraction based on 3D-FM is $( 0.032,0.034V)$. }
\label{fig9_n}
\end{figure}
The constructed ferromagnetic Hamiltonian model has three categories of dimension:1D, 2D, 3D. Fig.~\ref{fig8} displays three connection types for ferromagnetic model. All P-Bit states in ferromagnetic model exhibit a uniform orientation, and each P-Bit simultaneously interacts with $2\times Dimension$ adjacent P-Bits. For example, the P-Bit in 3D ferromagnetic model has 6 adjacent P-Bits. The coupling strength $J$ between P-Bits is the same, which means that every P-bit is equivalent. The right side gives the weight matrix and density matrix for ferromagnetic model. In the weight matrix, the total number of 1 in each row is: $2\times Dimension$, which corresponds to the number of adjacent P-Bits for each P-Bit. The 1 which situated on the last position of anti-diagonal in the matrix signifies the presence of periodic boundary conditions. The bias $B$ on the diagonal of matrix are all 0. For the calculated density matrix, the total number of 1 in each row is: $2\times Dimension+1$, since the density matrix has an order higher than the weight matrix, and the diagonal of density matrix is 1. The last column in matrix is 0, which infers that the ideal mean value of P-Bit state in ferromagnetic Hamiltonian model is 0.

The three types of dimension for ferromagnetic Hamiltonian model we have developed are meticulously designed to address the critical aspect centered on the variation extraction algorithm highlighted in the preceding analysis. However, ascertaining the dimensionality of the ferromagnetic model that most compatible with the extraction algorithm requires empirical verification. To initiate this validation process, we measured the standard deviation of the variation for 600 fabricated P-Bit devices. The standard deviation of $\alpha$ is approximately 0.3, and for $\Delta V$, it is around $1V$. Utilizing this level of P-Bit variation, we proceed to generate a substantial number of P-Bit devices with diverse variations in simulations. In these simulations, the $\alpha$ follows a Gaussian distribution with a mean value of 1 and a standard deviation of 0.3, while that of the $\Delta V$ are 0 and 1$V$.
\begin{table}[b]
\caption{\label{STD}%
On different dimensional ferromagnetic models, we measure the statistical standard deviation of the variation extraction algorithm. The 'w/o extraction' signifies the initial deviation predetermined in our simulations. The statistical standard deviation of the extraction algorithm is computed as follows: $\sigma_\alpha=\sigma(\alpha_{extr}-\alpha_{real})$, $\sigma_{\Delta V}=\sigma(\Delta V_{extr}-\Delta V_{real})$.
}
\begin{ruledtabular}
\begin{tabular}{ccccc}
\textrm{}&
\textrm{w/o extraction}&
\textrm{1D-FM}&
\textrm{2D-FM}&
\textrm{3D-FM}\\
\colrule
$\sigma_\alpha$ & 0.3 & 0.116 & 0.061 & 0.032\\
$\sigma_{\Delta V}$ & 1V & 0.143V & 0.063V & 0.034V\\
\end{tabular}
\end{ruledtabular}
\end{table}

As shown in Fig.~\ref{fig9}, based on these P-Bit devices, we construct three kinds of P-Bit arrays including 1000 devices respectively utilizing 1D, 2D, and 3D ferromagnetic model (FM) to perform variation extraction. In Fig.~\ref{fig9}, different colors correspond to different P-Bits, where the solid lines demonstrate the evolving trends of extracted $\alpha$ and $\Delta V$ with the number of iterations, while the dash lines represent their real values. For the 1D-FM, there is a moderate deviation between the extracted values of $\alpha$ and the real values, while the extraction direction is accurate. The extracted values of $\Delta V$ are relatively close to the real values. In the case of 2D-FM, the training outcomes for the difficult $\alpha$ are satisfactory, with the extracted variation being relatively consistent with the real value. Furthermore, the training result for $\Delta V$ is highly precise, nearly identical to the real value. With regard to 3D-FM, the learning results are further improved compared to 2D-FM. The extracted values of $\alpha$ are almost identical to the real value, which is remarkable for the challenging extraction of $\alpha$. As for $\Delta V$, the extracted values perfectly align with the real values.

We conducted a detailed analysis of the statistical standard deviation exhibited by the extraction algorithm across various dimensional ferromagnetic models. As illustrated in Table~\ref{STD}, the implementation of the extraction algorithm based on the ferromagnetic model notably mitigates device variation levels within extensive P-Bit arrays. Of particular significance is the 3D-FM model, which achieves a noteworthy reduction in the standard deviation of $\alpha$ to an impressively low 0.032 and $\Delta V$ to an exceptionally minimal 0.034V. The specific Sigmoid curves before and after extraction are further provided. As shown in the Fig.~\ref{fig9_n}, the distribution of curves without extraction is quite broad, whereas the extraction based on 3D-FM can reduce the variation level to a sufficiently low range. As shown the orange point at Fig.~\ref{fig2_mk}, this variation can achieve accurate Ising computing. After the above comparison, the final results indicate that the extraction algorithm based on 3D-FM can most accurately extract $\alpha$ and $\Delta V$.

\subsection{Discussions of the Automatic Extraction Algrithm}\label{sec3d}
Upon scrutinizing the variation extraction outcomes performed on P-Bit array in the aforementioned models, a noteworthy observation emerges. Specifically, the extraction challenge associated with $\alpha$ is markedly more formidable than that associated with $\Delta V$. This discrepancy can be attributed to the inherent nature of their relationships with input voltage $V$, as delineated in Eq.~\ref{beha2}. It becomes evident that $\alpha$ exhibits a multiplicative dependence on $V_{input}$, while $\Delta V$ manifests an additive relationship with the same variable. Delving into the intricacies from the perspective of the weight matrix, the $\Delta V_i$ singularly impacts the $B_i$ corresponding to P-Bit $i$, whereas $\alpha_i$ exerts influence across all the $W_{ij}$ in the $i_{th}$ row of the weight matrix. This distinction in the impact scope between $\alpha$ and $\Delta V$ underlies the observed difference in extraction difficulty. Given the direct influence of $\alpha$ extraction on $\Delta V$ in the training process, achieving precise $\alpha$ extraction becomes crucial.

However, within weight matrix of the AND gate, the inconsistency of $W_{ij}$ results in a non-uniform influence of $\alpha$ on weights. This factor collectively affects the training of $\alpha$. Challenges emerge due to the inherent extraction inaccuracies of $\alpha$ and the non-scalable nature ingrained within the AND gate. Although addressing the accuracy of $\alpha$ training within the AND gate is achievable, achieving seamless scalability presents challenges. Cascading logic gates amplify the complexity of the weight matrix, rendering the training process notably intricate.

Building upon the aforementioned analysis, we have devised the ferromagnetic model characterized by uniform $W$, $B$, and inherent scalability. Through the training of 1D, 2D, and 3D ferromagnetic models, a consistent improvement in training effectiveness is observed with the increasing dimensionality, particularly notable in the case of 3D-FM. In this scenario, the proposed extraction algorithm simultaneously and adeptly captures all variations in P-Bits.

The enhanced performance in higher-dimensional ferromagnetic models can be attributed to their increased stability. According to the Landau theorem, 1D ferromagnetic materials are significantly influenced by thermal fluctuations, thus making it challenging for them to exhibit ferromagnetism. The 2D ferromagnetic materials lose their magnetic properties beyond the Curie temperature. In contrast, the 3D ferromagnetic materials possess higher degrees of freedom, experiencing less influence from entropy effects, thus exhibiting greater stability. In a more stable 3D ferromagnetic Hamiltonian model, the extraction of $\alpha$ becomes notably clearer and more accurate.

It is worth noting that in a digital system involving both read and write processes, our method may not be more efficient than direct exponential fitting. However, in an analog system, the situation changes because convenient read and write operations are absent. Moreover, digital systems have two significant drawbacks compared to analog systems: 1) Power consumption: Digital systems using FPGA and DAC consume considerable power. 2) Speed: Digital systems require clocks for read and write operations, which means that the flipping speed of P-Bit devices is limited by the FPGA clock. The speed advantage of stochastic MTJ (ns) might be compromised by the read operations in digital systems (see supplementary materials in~\cite{Yang1sup}). In contrast, an analog Ising system composed of resistors/capacitances networks~\cite{analog1,analog2} or tunable logic gates will have lower energy consumption as it does not require application-specific integrated circuits (ASICs) and high-precision DACs. Regarding speed, the advantage of ultra-fast stochastic MTJs~\cite{Ohno1} will be realized because the spins in an analog system can update asynchronously. The local minima in Ising computations will be mitigated by the speed differences between MTJs~\cite{IEDM1,Zeng1}.
However, analog Ising systems face calibration challenges since the Sigmoid curve of P-Bits cannot be acquired directly. Therefore, our method is designed to be more suitable for analog Ising systems, where read and write operations are not feasible, and direct exponential fitting cannot be performed. In such cases, our method helps extract the individual variations of each device and uses a weight compensation algorithm for correction, thereby achieving accurate results in Ising computing.
\section{Transferable and Scalable Ising computations on Large P-Bit Array with Variations}\label{sec4}
In addressing the challenge of device-to-device variations on P-Bit array within Ising computations, our proposed approach involves the utilization of both a weight matrix compensation method and an variation extraction algorithm. A crucial sequence for practical implementation becomes apparent: Extraction and Compensation. Specifically, for a substantial P-Bit array with variation, we initially employ the proposed extraction algorithm to extract the variations of each P-Bit, and store them for later use. Owing to the simplicity and scalability of the 3D ferromagnetic model employed in the extraction algorithm, the extraction based P-Bit array can be greatly expanded. Given the invariant Sigmoid curve of P-Bit devices, their $\alpha$ and $\Delta V$ remain constant. In essence, after performing the extraction algorithm once, we can consider the variations as inherent attributes of the P-Bits. Subsequently, we can utilize these variations in conjunction with weight matrix compensation to tackle diverse combinatorial optimization problems. Due to the transferability inherent in the weight compensation method, repetitive training is unnecessary. By incorporating specific compensation parameters denoted as $C_\alpha$ and $C_{\Delta V}$ into weight matrix, the real P-Bits with variations can be treated as ideal P-Bits without variations for computational purposes. Hence, we can achieve transferable and scalable Ising computation using a large P-Bit array with variation.
\begin{figure*}[!tb]
\centerline{\includegraphics[width=2.1\columnwidth]{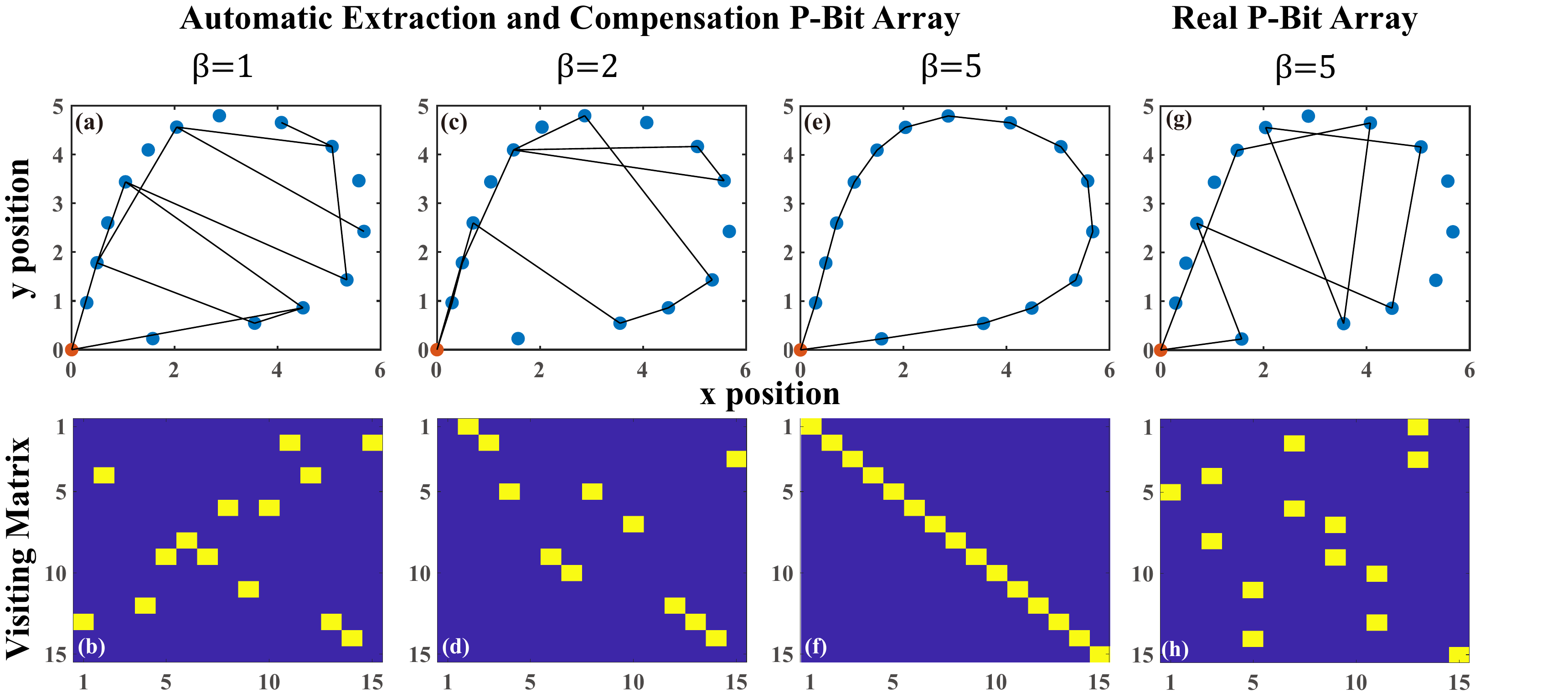}}
\caption{The computing results of 16-city TSP solved by both the Automatic Extraction and Compensation P-Bit array and the real P-Bit array with variations. (a), (c), (e) The route of the 16-city TSP from the annealing process at $\beta=1, \beta=2, \beta=5 $ based on Automatic Extraction and Compensation array. As $\beta$ gradually increases to a sufficient level, the Ising system ultimately computes the shortest and compliant tour length. For comparison, we observed that at $beta=5$, the real array failed to yield the correct results. (b), (d), (f), (h) Their respective visiting matrix $m$. }
\label{fig10}
\end{figure*}
\begin{figure}[!tb]
\centerline{\includegraphics[width=0.9\columnwidth]{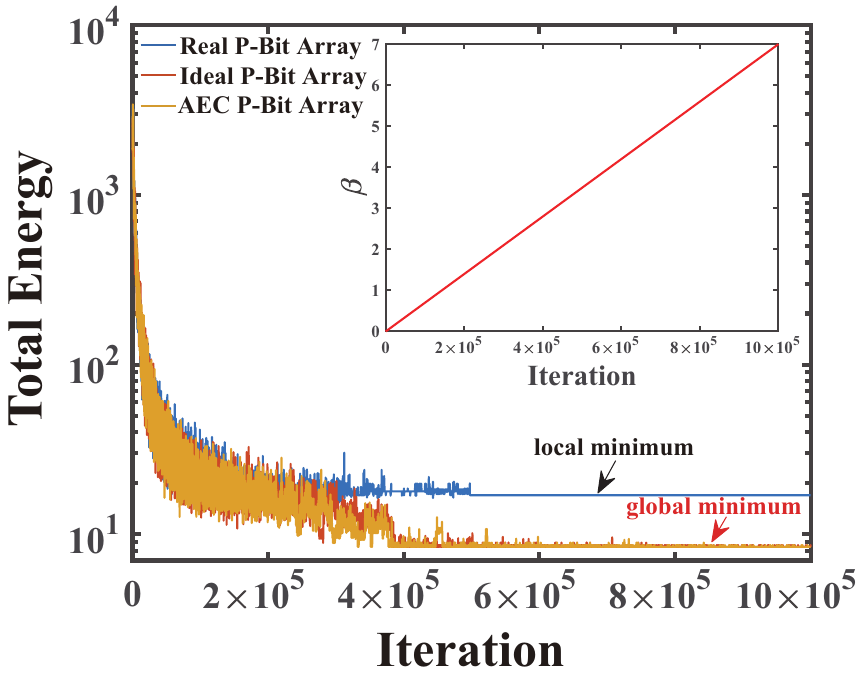}}
\caption{The total energy of annealing process for 3 situations of P-Bit array. The system of real P-Bit array with variation is trapped in a local minimum, while the ideal P-Bit array and the real P-Bit array with Automatic Extraction and Compensation algorithm can facilitate the systems annealing to the global minimum.}
\label{fig11}
\end{figure}
\subsection{Successful Solution for 16-city Traveling Salesman Problem}\label{sec4a}
To validate the feasibility, transferability, and scalability of our proposed extraction algorithm and weight compensation method, we employed a large-scale P-Bit array to solve the traveling salesman problem (TSP). It is a challenging NP-complete problem that aims on finding the shortest path to visit $N$ cities. The Ising Hamiltonian of TSP can be written as follow~\cite{Datta2}:
\begin{eqnarray}
H_{TSP}=\sum_{v=1}^{N}{(1-\sum_{j=1}^{N}{m_{v,j}})^2}+\sum_{j=1}^{N}{(1-\sum_{v=1}^{N}{m_{v,j}})^2}\nonumber\\
+\lambda_p\sum_{uvj}W_{(uv)}m_{u,j}m_{v,j+1}
\label{TSP_equ}
\end{eqnarray}
where $m$ is the visiting matrix, $m_{v,j}$ means that, at the  $j_{th}$ time, the selection is made to visit city $v$. The first constraint indicates that a city can be visited only once, the second constraint implies that each visit should involve only a single city. The third item concerns the calculation of the travel path length, with $\lambda_p$ serving as the penalty coefficient, $W_{{uv}}$ is the distance between city $u$ and city $v$. It is worth noting that this constraint should take into account the distance between the starting city and the first chosen city, as well as the distance between the last visited city and the starting city, given that the TSP requires returning to the starting point eventually.
\begin{table}[b]
\caption{\label{TSP}%
The average path energy of 100 trials for TSP16 across 3 P-Bit arrays. The average path energy from the Automatic Extraction and Compensation P-Bit array can be brought very close to that of the ideal array.
}
\begin{ruledtabular}
\begin{tabular}{cccc}
\textrm{P-Bit Array} & \textrm{Ideal} & \textrm{Real} & \multicolumn{1}{c}{\parbox[c]{3.6cm}{\centering \textrm{Automatic Extraction and Compensation}}} \\
\colrule
\multicolumn{1}{c}{\parbox[c]{2.4cm}{\centering \textrm{Average Path Energy}}} & 17.837 & 34.080 & 18.080\\
\end{tabular}
\end{ruledtabular}
\end{table}

For a N-city TSP with given starting point, calculating the optimal tourist route requires $(N-1)^2$ P-Bits. As illustrated in Fig.~\ref{fig10}, we compute a TSP involving 16 cities. Based on the P-Bit array with variations generated in simulations (see Sec.~\ref{sec3b}), we construct a 3D ferromagnetic Hamiltonian model, then extract the variation of each P-Bit from the 225-bit array. By employing the weight compensation method, we can mitigate the device-to-device variations in the computation. With an annealing technique, we finally obtain the correct route.

In the Fig.~\ref{fig10}(a), (c), (e) and (g), the $\beta$ represents the inverse temperature. The red point means the starting city is (0,0). In observing the Automatic Extraction and Compensation P-Bit array, it becomes apparent that, with the progression of annealing , the increment in $\beta$ gradually directs the salesman to visit each city while progressively reducing the total tour length. Eventually, as the $\beta$ approaches 5, the Ising system converges to find the shortest and compliant tour length. For comparison, the Fig.~\ref{fig10}(g) illustrates the inability of the real array with variations to produce the correct outcome when $\beta$ is set to 5. This highlights the efficacy and precision of the our proposed Automatic Extraction and Compensation algorithm. The Fig.~\ref{fig10}(b), (d), (f) and (h) displays the respective visiting matrix $m$. Fig.~\ref{fig11} shows the evolution of the total energy during the annealing process performed by 3 types of P-Bit array: real, ideal and Automatic Extraction and Compensation, where the insert figure displays the trend of $\beta$ with the epochs. We note that the variations present in the real P-Bit array tend to stagnate the Ising system in a local minimum. Nonetheless, the ideal P-Bit array and the real P-Bit array with Automatic Extraction and Compensation algorithm can facilitate the system's annealing to the global minimum.

We further evaluated the average path energy of TSP-16 over 100 trials across 3 different P-Bit arrays. As shown in Table.~\ref{TSP}, the average path energy differs markedly before and after applying our automatic correction method. The average path energy from the P-Bit array using the Automatic Extraction and Compensation method can be reduced to nearly match that of the ideal array.

\subsection{Successful Solution for 21-bit Integer Factorization}\label{sec4b}
\begin{figure*}[!tb]
\centerline{\includegraphics[width=1.8\columnwidth]{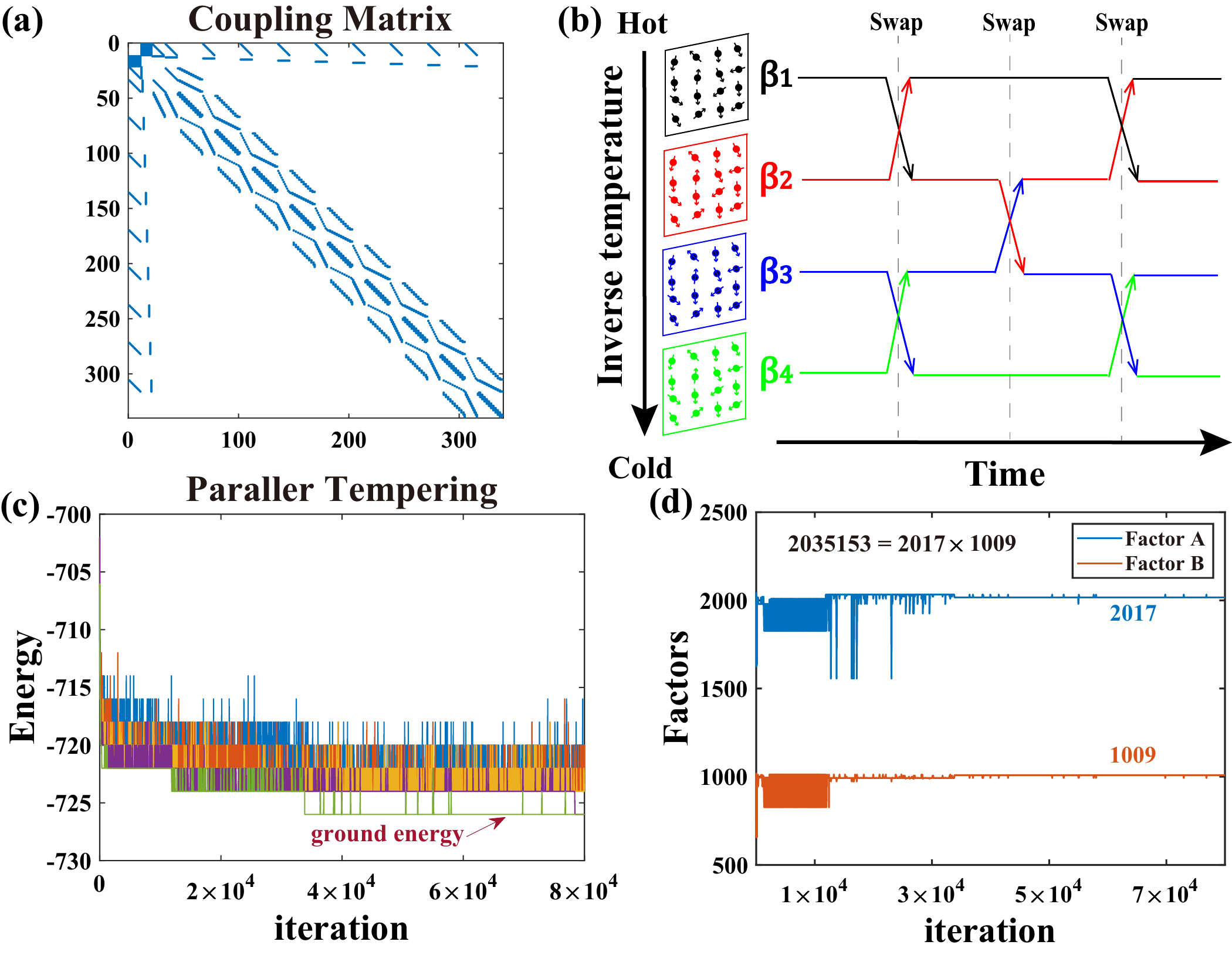}}
\caption{(a)The coupling matrix for the factorization of a 21-bit semi-prime. (b)The illustrator for Parallel Tempering. (c)The energy trends of the replicas at different $\beta$. (d)The trends of factors A and B with increasing iterations.}
\label{fig12}
\end{figure*}
To demonstrate the effectiveness of our proposed algorithm across multiple problems, we further performed a 21-bit semi-prime factorization calculation on a large P-Bit array with variations, successfully factoring 2035153. 
As shown in Fig.~\ref{fig12}, our proposed Automatic Extraction and weight Compensation algorithm can reduce the variation of the P-Bit device array sufficiently to complete the challenging task of semi-prime factorization. This factorization was achieved using invertible logic method~\cite{Datta3,Datta6}, which clamps the output of the multiplier to perform semi-prime factorization. According to the invertible logic method, Fig.~\ref{fig12}(a) presents the coupling matrix for the factorization of a 21-bit semi-prime 2035153. Additionally, we employed the Parallel Tempering (PT) method~\cite{PT1,PT2}, generating multiple replicas at different temperatures to search for the optimal state. As shown in Fig.~\ref{fig12}(b), periodically, these replicas were swapped. These swap stages enable the whole Ising system to escape local optima, ensuring that the lowest energy state operates at the lowest temperature (highest $\beta$). The swap probability between two replicas is written as~\cite{PT3}:
\begin{equation}
P_{swap}=min\{1,exp(\Delta\beta(E_i-E_j))\}
\label{PT_swap}
\end{equation}
where $E_i$,$E_j$ are the energies at inverse temperature  $\beta_i$, $\beta_j$. When the energy $E$ is higher at a higher $\beta$, a lower energy state will be exchanged at this $\beta$.
Fig.~\ref{fig12}(c) shows the energy trends of the replicas at different $\beta$, where different colors represent the energy states at different $\beta$ values. The Ising system continuously searches and swaps low-energy states to the higher $\beta$, ultimately ensuring that the lowest energy state is maintained at the highest $\beta$. Fig.~\ref{fig12}(d) presents the values of factors A and B during the PT process. After approximately 35000 iterations, the correct factorization was achieved: $2035153=2017\times1009$.

The effectiveness of our algorithm has been validated in both the challenging 21-bit semi-prime factorization and the TSP-16 problem performed on a large P-Bit array with variations. It has been demonstrated that the Automatic Extraction and Compensation algorithm can reduce the variations in the P-Bit array to a sufficiently low level to accurately perform complex computations.

The obtained results validate the efficacy of the proposed extraction algorithm and weight compensation method. The individual calibration needs of P-Bit devices are addressed through weight compensation. Due to the transferability inherent in this method, the weight matrix rederivation enables the treatment of real P-Bits as ideal ones for diverse problems, eliminating the necessity for weight matrix retraining. Furthermore, the extraction algorithm, coupled with 3D-FM, accurately extracts all variations within the P-Bit array, demonstrating scalability across array sizes. By clearly following an Extraction-Compensation sequence, a transferable (for any problem) and scalable (for any array size) Ising computation on the P-Bit array with variations is facilitated.
\section{Conclusion}\label{sec5}
In this work, we initially proposed the behavioral model incorporating $\alpha$ and $\Delta V$ which fits our fabricated SOT P-Bit devices well. Then we introduce the Automatic Extraction and Compensation algorithm. The Extraction step can extract the variations of all the P-Bit devices across the whole large P-Bit array simultaneously, while the Compensation step mitigates the device variations by modifying the weight matrix according to the extracted variations. Finally, we employ the developed Extraction-Compensation process to facilitate accurate solutions for both 16-city TSP spanning 225 bits and 21-bit semi-prime factorization, thereby confirming the efficiency, transferability and scalability of our algorithm. The term 'efficiency' indicates that individual calibration for P-Bits with variations is not required. The term 'transferability' means that weight matrix retraining for diverse problems is unnecessary. 'Scalability' refers to the capability of expanding the array scale. The Automatic Extraction and Compensation algorithm enables accurate and efficient completion of Ising computations on large P-Bit arrays with variations, therefore improving the efficiency, accessibility, and widespread applicability of probabilistic computers.

\begin{acknowledgments}
The authors wish to acknowledge the support from the National Natural Science Foundation of China under Grant 62171009, T2293703, T2293700, 61901017, 62172025, U2241213, Beijing Natural Science Foundation under Grant 4232069, Zhejiang Provincial Natural Science Foundation of China under Grant LZ24F040003 and Postdoctoral Science Foundation of China under Grant 2022M720368.

B.L.Z. and Y.L. contribute equally to this work.
\end{acknowledgments}
\bibliography{Auto_correction}

\end{document}